
\input phyzzx

\overfullrule=0pt

\def\Gv{G_{vac}(1,2)}

\tolerance=5000
\sequentialequations
\overfullrule=0pt
\twelvepoint
\nopubblock
\PHYSREV
\line{\hfill IASSNS-HEP-93/12}
\line{\hfill February 1993}
\line{\hfill hep-th/9302096}
\titlepage
\title{Quantum Purity at a Small Price: Easing a Black Hole Paradox}
\author{Frank Wilczek\foot{Research supported in part by DOE
grant
DE-FG02-90ER40542}
\foot {Not
a talk given at the Houston Conference on Black Holes, January 1992.
(See text). To appear in the Proceedings.}}
\vskip.2cm
\centerline{{\it School of Natural Sciences}}
\centerline{{\it Institute for Advanced Study}}
\centerline{{\it Olden Lane}}
\centerline{{\it Princeton, N.J. 08540}}
\endpage

\abstract{Following Hawking, it is usual to mimic the effect of
collapse space-time geometry on quantum fields in a semi-classical
approximation by imposing suitable boundary conditions at the origin of
coordinates, which effectively becomes a moving mirror. Suitable mirror
trajectories induces a close analogue to the radiance of black holes,
including a flux of outgoing radiation that appears accurately thermal.
If the acceleration of the mirror eventually ceases the complete state
of the radiation field is a pure quantum state, even though it is
indistinguishable from an accurately thermal state for an arbitrarily
long period of time and in a precise sense differs little from
``pure thermal'' closely followed by ``vacuum''.    Suspicions that the
semiclassical calculation of black hole radiance gives evidence for the
evolution of pure into mixed states are criticized on this basis.
Possible extensions of the model to mimic black holes more accurately
(including the effects of back reaction and partial transparency),
while remaining within the realm of tractable models, are suggested.}

\endpage

\REF\bhhair{S. Coleman, J. Preskill, and F. Wilczek, {\it Nucl. Phys.} {\bf
B370}, 577 (1992).}

\REF\pert{C. Holzhey and F. Wilczek, {\it Nucl. Phys.} {\bf B380}, 447
(1992).}

\REF\thesis{C. Holzhey, Princeton University Thesis, (1992),
unpublished.}

\REF\infpar{S. Hawking {\it Comm. Math. Phys.} {\bf 43}, 199, (1975).}

\REF\birrell{N. D. Birrell and P. C. W. Davie, {\it Quantum Fields in
Curved Space}, Cambridge University Press, Cambridge (1982).}

\REF\hawk{S. Hawking, {\it Phys. Rev.} {\bf D14}, 2460, (1976).}

\REF\carlwil{R. Carlitz and R. Willey {\it Phys. Rev.} {\bf D36}, 2327,
2336, (1987).}

\REF\unruh{W. Unruh, {\it Phys. Rev} {\bf D14}, 870, (1976).}

\REF\mis{C. Misner, K. Thorne and J. Wheeler, {\it Gravitation},
Freeman, NY (1973).}

(The talk as given reported on work that has now been
published [\bhhair ] [\pert ]
and I see no point in duplicating this material.
Instead I am taking the
opportunity to discuss some joint
work contained in my student
Christoph Holzhey's thesis [\thesis ].  The discussion here
is qualitative.  More quantitative discussions
including comparison between coarse and fine-grained
entropy for particular cases of radiance are contained in
the thesis, and
we plan to write them
up fully in due course.)
\bigskip
\bigskip
\bigskip

Semi-classical calculations of the radiation [\infpar, \birrell]
 from black holes
indicate that
their emission is the same as one might expect from an ideal gray body.
It is commonly believed that these calculations are very accurate for
black holes having masses much larger than the Planck mass (and away
from any extremal limit).   This raises a conceptual problem that has
been much discussed, as follows.
One can certainly imagine forming a black hole from
matter in a pure quantum state.  One then finds, in an
approximation which appears accurate,
that it radiates to produce a mixed
state.  Yet the evolution of a pure into a mixed state
would violate the basic
principles of quantum mechanics as they are currently understood.
Inspired by this conflict,
Hawking [\hawk] boldly proposed to jettison some of the
usual assumptions of quantum mechanics.  However,
naturally  many physicists are
reluctant to tinker with
the foundations of our most practically successful
physical theory in response to such an esoteric thought experiment, and
so far this approach has not been very fruitful.

Indeed there are many things we do not understand about quantum gravity,
and one's first reaction might be to hope that this conceptual
problem will resolve itself once we obtain a proper understanding of
the ultraviolet properties of gravity,
in particular its behavior under conditions of extreme curvature.
According to the classical theory of black holes,
such regions -- singularities -- do inevitably occur.  They are
repositories for our ignorance, and appear to provide  promising
scapegoats for any conceptual problems we encounter with black holes.

After
a little further reflection, however, it may come to seem
less reasonable to hope that
any possible
behavior near the singularities could help  resolve the
particular puzzles posed by black hole radiance.
For the bulk of the predicted radiation
from a massive black hole can be demonstrated
to arise in an analysis where large curvatures
never appear, and quantum corrections due to fluctuations of
the gravitational field ought to be negligible.  (Of course because
of ultraviolet problems the graviton loop corrections, not only
to this but to all physical processes, diverge.  However any ``reasonable''
cutoff procedure, such as supposing that the Planck mass sets
the scale for the magnitude of the coefficients of
higher-dimensional operators
in the effective Lagrangian, leads one to expect a small result.)
In other words, the semi-classical
approach to the radiance, leading to apparently thermal radiation,
governs the bulk of the history of the black hole evaporation process,
as measured
for example by the energy or the entropy of the emitted radiation.
Thus one might be tempted to say -- and it has often, in the
literature, been stated -- that by the time the black hole gets to the
Planck mass and quantum gravity effects plausibly come into play  it is too
late.  For at that point
there is no longer enough material (energy) left to restore the missing
correlations, which are necessary to render a mixed state with arbitrarily
gigantic entropy pure.

Here it is demonstrated, in the context
a simple soluble model
that can mimic many of the features of black hole radiance, that
this argument is too quick.
The model in question is the imposition of boundary conditions on
a moving surface.  In the simplest case one imposes reflection boundary
conditions, thus arriving at the {\it moving mirror\/}  model.  The global
structure of the Cauchy problem in this model is quite transparent, and
one can address the question of quantum purity on the basis of general
principles.  On the other hand the moving mirror boundary conditions
generally induce radiation, that for suitable trajectories can closely
mimic the Hawking radiation from black holes.  We shall see that
quantum purity is enforced in subtle ways at little cost in
energy, even after an arbitrarily long period of apparently
thermal radiation.  Very roughly, the result of
these considerations is to put the paradox back where it belongs, at the
Planck scale -- and also, as we shall see, to suggest
plausible
model pictures of how it might be resolved.

Of course the moving mirror
model is not new (see the references in [\birrell]).
Indeed it was used with a philosophy  similar to
the one adopted here in two very ingenious papers by Carlitz and
Willey [\carlwil ],
which appear not to have received the attention they deserve.

\chapter{Causal Structure of the Mirror Problem}

Consider the evolution of a massless scalar field
in 1+1 dimensions subject to the
boundary condition
$$
\phi (z(t), t) ~=~ 0
\eqn\globa
$$
along the mirror trajectory $z = z(t)$.
The scalar field is defined to vanish on the left-hand side of the
mirror.
The effect of the boundary condition is of course that rays incident
on the mirror reflect off it.

As will be discussed below,
the mirror plays the role of the origin $r=0$ in
space -- the center of the hole -- in the black hole problem.
Thus reflection off the mirror mimics the propagation of an ingoing
wave to the center and its emergence as an outgoing wave.  The
distortion of space-time -- essentially the
lengthening of space (and shortening of time) near the surface of
the hole --
in during collapse has a dynamical effect similar
to
the
effect of a
{\it rapidly receding\/} mirror in 1+1 dimensional flat space.
Indeed the fundamental effect
is that rays reflected off a rapidly
receding mirror are severely red-shifted --
as are the rays, crucial to Hawking's analysis, which
barely avoid being trapped behind the incipient horizon.

Three types of mirror trajectories
are illustrated in Figures 1-3.

The first trajectory type describes a mirror
that accelerates away
from rest at
$t=0$ and approaches the speed of light asymptotically.
Let the asymptote light-ray be denoted $A$ as in Figure 1.
Since we
are dealing, for simplicity,
with a massless field we may consider only left-moving modes.  Let us
define points 1, 2, 4, 5 as in the Figure, and use the same labels to
distinguish
the rays emanating from these points
at $t=0$.

We see that rays such as 1 and 2, which begin to the left of $A$, intersect
the mirror and propagate out to spatial infinity at the right, denoted
in deference to the black hole interpretation as ${\cal I}_+$.  On the
other hand rays such as 4 and 5, which begin to the right of $A$, never
intersect the mirror.  They propagate to the left infinity, denoted as
${\cal H}_+$.  (This infinity may seem  a little funny from the
point of view of the
metaphorical interpretation of
$z$ as the effective position of the black-hole origin.  The point is
that the effective radial distance from the point of view of wave
propagation is most
appropriately measured in intervals of the tortoise coordinate
$r_*$, which diverges to $-\infty$ at the black hole horizon.
By the way, these rays leave the Figure in finite affine time.)

Now consider the problem of the the evolution of a quantum state
defined for $z>0$ at $t=0$ into the distant future.  Naturally one should
consider first the ground state, defined by the absence of
positive-frequency modes.
It is evident that the time interval between
the arrival of 1 and 2 at a given point in space before they
reflect is much dilated after they reflect.  Thus the frequency of
waves is altered, and negative-frequency waves can acquire positive-frequency
components.  This would be interpreted as the creation of an excited
state on ${\cal I}_+$.  For an appropriate mirror trajectory, as
we shall see,
the state on ${\cal I}_+$ will be a thermal state, with its temperature
related to the rate of acceleration of the mirror.
Clearly all information concerning the state of the field
$\phi$ in region to the left of $A$ at $t=0$ is propagated to
${\cal I}_+$.

Rays such as 4 and 5 beginning to
the right of $A$ propagate undisturbed to
${\cal H}_+$.
Clearly all information concerning the state of the field
$\phi$ in region to the right of $A$ at $t=0$ is propagated to
${\cal H}_+$.
If we start with the ground state
on $t=0$, an
observer making measurements on ${\cal H}_+$
also sees his natural
ground state.

Now according to basic principles of quantum mechanics, which of course
are certainly
not contradicted by anything in the simple
model problem under consideration,
a pure state localized to the left of $A$ would propagate into a pure state on
${\cal I}_+$ and a pure state localized to the right of $A$
 would propagate into a pure
state on ${\cal H}_+$.  However, the ground state at $t=0$ is
{\it not\/} pure when restricted either to either segment.
The positive-frequency
condition forces consideration of modes which
extend over both intervals, and introduces
correlations between these intervals.   Indeed, the two-point function
$\langle \phi (1) \phi (4) \rangle $ at $t=0$, for example,
certainly does not vanish.  Furthermore, this correlation will
propagate in a simple way into the future, introducing correlations
between ${\cal I}_+$ and ${\cal H}_+$.
Thus we should not be shocked to find
a mixed state if we consider ${\cal I}_+$ by itself, without
regard to (tracing over) the state on ${\cal H}_+$.   And this indeed is
what we do find: the correlation functions on ${\cal I}_+$, for the
appropriate trajectory of this type, are {\it precisely\/} thermal,
and therefore
certainly must be described by a mixed state on ${\cal I}_+$.

The phenomenon that may be
a shock to one's intuition
is that
{\it it is correlations between the rich thermal state on
${\cal I}_+$ and the
apparently barren desert on ${\cal H}_+$
which insure purity of the whole}.  Thus for example the
expectation value of the energy-momentum tensor vanishes,
and its multi-point correlators are vacuous (\ie\ indistinguishable
from the vacuum),
when restricted to
${\cal H}_+$ -- but its cross-correlators between ${\cal H}_+$ and
${\cal I}_+$ do not vanish.
This peculiar phenomenon, whose existence and nature is made quite
transparent by the foregoing extremely elementary observations,
was noted and emphasized by
Carlitz and Willey.
(However they somewhat obscured the issue by claiming
in effect
that particle creation on ${\cal I}_+$ is uniquely and locally
related
to particle creation on ${\cal H}_+$, which is not
the case.)
It shows in as dramatic fashion as one could desire
that the purity of a big complicated state with gigantic entropy
(in any sense) can be restored at little -- here
actually at {\it zero\/}
-- cost in energy.

Now let us consider the mirror trajectory
depicted in Figure 2, which is the same as the one discussed
for a long interval of time, but such that the mirror eventually
stops accelerating.  Then all rays eventually intersect the
mirror, and get reflected to ${\cal I}_+$.  Thus we obtain on
${\cal I}_+$ a pure state which looks thermal for
an arbitrarily long time.  Of course once the mirror stops accelerating
there is no longer any radiation emitted.  The transition to
zero acceleration can be done smoothly, so that only a small burst
(whose magnitude is essentially independent of the
length of the interval over
which thermal radiation has occurred)
accompanies it.  Thus altogether one finds, similar to the
previous case, that quantum purity comes at a small price.

The situation of Figure 2 is not yet a
model for
complete black hole evaporation.  For although
positive frequencies at late times are indeed reflected into positive
frequencies, and there is no particle production, yet the frequency
is highly red-shifted.  Thus real particles at late times will sense
(in the interpretation of the mirror as the locus of the
origin) a remnant that
delays them for a long time and saps their energy.  It is
left for the reader
to invent witty names for such a remnant.

Finally in Figure 3 we have the situation where the mirror returns
to rest.
Real particles emitted at late times, which intersect
the mirror during its second period of rest, behave as if passing through
the origin of empty space in the analogue problem.
Thus this provides a model for a black hole that evaporates completely.
{}From the nature of the construction, any pure initial state evolves into
a pure final state.

\chapter{Connection to Collapse Geometry and Radiance}

Now I would like to indicate, following
Unruh [\unruh ] a precise connection between
the mirror model and collapse geometry, and summarize
the resulting
correlation functions and radiation flux.

Consider for simplicity
a spherically symmetric collapsing shell of matter.
We have
vacuum inside and outside the shell, while the shell carries a given amount
of mass (and possibly other quantum numbers).
Thanks to Birkhoff's theorem, we know the
metric in both  regions:
$$ds^2= \cases{dr^2-d \tau^2-r^2 d\Omega^2,
                & for $\tau+r \le V_s $;\cr
        \lambda^2 dt^2-\lambda^{-2} dr^2 -r^2 d \Omega^2,
                & for $t+r \ge v_s$. } \eqn\meta$$
Note that in order to exhibit the metric in each region in its familiar
(static) form, two different sets of coordinates had to be used.   It is
convenient to introduce light-cone coordinates in each region. In
the interior region we use simply
$U=\tau-r$ and $V=\tau+r$, whereas in the outer region we first define the
tortoise-coordinate $r_*$ through
 $${dr_* \over dr}= {1 \over \lambda^2}, \eqn\rtortdef $$  and then take
$u=t-r_*$ and $v=t+r_*$ as light-cone coordinates. The  space-time is
described by the metric:
$$ds^2=\cases{dUdV-r^2 d\Omega^2, & for $V\le V_s$; \cr
        \lambda^2 dudv-r^2 d\Omega^2, & for $v\ge v_s$ ,}
        \eqn\metb$$
where $r$ is determined through the relations
$$\eqalign{V-U=2 r, \qquad & {\rm for }\quad  V\le V_s; \cr
                v-u=2 r_*(r), \qquad & {\rm for }\quad  v\ge v_s.}
         \eqn\rofuv $$

When we paste together the two coordinate systems for the interior and
exterior region to form a global coordinate-system, we can choose to
coincide with \metb\ either
in the exterior or in the interior region. The first choice
is natural from the point of view of a distant observer,  while the second
is more convenient to implement the boundary condition at the
origin and to display the complete space-time structure.

Let us consider first the
former choice, that is
using $u$-$v$-coordinates in both regions and looking for a
satisfactory coordinate-transformation $U(u)$
and $V(v)$. In the infinite past the space-time is flat and there is no
difference between the two coordinate systems. This implies that we can
choose $V(v)=v$. We find the function $U(u)$ by demanding that along
the worldline $v=v_s$ of the shell the coordinate $r$ should agree in both
systems, because it has a gauge-invariant meaning (it determines the area
of a two-sphere at constant radius and time). Applying \rofuv\  along
$v=v_s$ we  obtain the  implicit relation:
$$r_*\left(r={v_s-U(u)\over 2} \right)= {v_s-u \over 2}. \eqn\Uofu $$
Differentiating  this equation along the worldline of the shell we find,
with the help of the defining equation \rtortdef\ for $r_*$,
$${dU \over du}=\lambda^2(u,v_s),
        \eqn\dUdu$$
so that the metric becomes:
$$ds^2=\cases{\lambda^2(u,v_s) dudv-r^2 d\Omega^2,
         & for $v<v_s$; \cr
        \lambda^2(u,v) dudv-r^2 d\Omega^2, & for $v>v_s,$ }
        \eqn\metuv$$
which is continuous along $v_s$. The metric is, of course, only valid for
non-negative values of $r$, \ie\ for $v \ge U(u)$. The world-line of the
origin is therefore described by
$$v_o(u)=U(u). \eqn\origineq $$
Since nothing can go beyond the regular origin, \ie\ to negative $r$, it
acts like a perfectly reflecting mirror.

In the $u$-$v$-frame the shell never crosses the horizon since $r_*$
and $t=v_s-r_*$ diverge as the horizon is approached. On the other
hand we know that the shell reaches the origin in finite proper time.  In
order to describe the whole space-time, including the interior of the black
hole
it is convenient to use the
$U$-$V$-coordinates, which provide a complete cover since they contain
the  origin until the shell reaches it. The space-time is then described by
$$ds^2=\cases{ dUdV-r^2 d\Omega^2,
         & for $v\le v_s$; \cr
        \lambda^2(u,v) \lambda^{-2}(u,v_s) dUdV-r^2 d\Omega^2,
         & for $v\ge v_s,$}
        \eqn\metUV$$
In spite of its appearance, the metric is regular on the horizon where
$\lambda^2=0$. The origin is stationary at $V=U$ until the shell reaches it.

For a shell of mass $M$ one has explicitly for
the tortoise coordinate
$$r_*(r)=r+2 M \ln |r-2 M|+c,
        \eqn\rtorts$$
and thus from \Uofu
$$u=U-4M \ln\left|(-4M-U+v_s)/2\right|-2 c. \eqn\uschwarz $$
$c$ is here an arbitrary integration constant. As $U$ approaches
$U_h=v_s-4 M$, $u$ diverges, which identifies  the line $U=U_h$
with the future horizon.  Alternatively, the finite range of $U$ implies
according to \origineq\ that the origin approaches the light-like asymptote
$v=U_h$ at late times as viewed in the $u$-$v$-frame. At very early times,
the origin is at rest because as $u\to -\infty$,
$U \approx u$.  At late times we can invert
\uschwarz\  by neglecting the linear term. We  find that $U(u)$
is of the general form
$$U(u) = c_1+c_2 e^{- \kappa u},  \eqn\trath$$
where $\kappa=1/4M$ is the surface gravity.  (This relation is nothing but
the familiar transformation [\mis] between Eddington-Finkelstein
and Kruskal-coordinates:
$$U_K =-4 M e^{-u/4M}.
        \eqn\Ukrus$$
At late times our coordinate $U$ therefore agrees with Kruskal $U_K$,
while $V$ equals $v$  is always of the Eddington-Finkelstein type.)

The upshot of all this is simply to justify partially
but precisely the idea expressed earlier,
that the collapse geometry may be modeled by a moving mirror problem.
In this model the mirror arises at the origin of coordinates
({\it not\/} the horizon); its ``motion'' is an effective representation
of the distortion of space-time in the collapse.  An important feature
left out of the model in its simplest
form is the non-trivial spatial curvature outside the
shell.

For simple forms of matter, \eg\ a free massless scalar field, the
moving mirror problem is eminently tractable.  Any quantity of interest
may be calculated explicitly.  For example one has for the
correlation function of the fields
$$\eqalign{\Gv \equiv \bra 0 \phi(1) \phi(2) \ket 0 &= \cr
        ={1\over 4 \pi} \ln
        {(U_2-U_1+i \delta) (V_2-V_1+i \delta) \over
        (U_2-V_1+i \delta)(V_2-U_1+i \delta)} .\cr }
\eqn\gvacb $$
Indeed this function manifestly satisfies the wave equation with
the correct singularity and the moving mirror boundary condition, and
reduces to the correct vacuum value before the mirror motion begins.
For the energy-momentum tensor describing the emitted radiation one
finds
(see, for example, [\birrell])
$$\bra 0 T_{\mu \nu} \ket 0 ={\delta_{\mu u} \delta_{\nu u}
          \over 12 \pi} \sqrt{U'} {d^2 \over du^2} \sqrt{1/ U'}.
         \eqn\tmunu$$

All energy $n$-point functions are
determined by $\bra 0 T_{\mu \nu} \ket 0 $ and $\Gv$.  For example
the energy two-point function
$$C_{\mu\nu ,\alpha\beta}(1,2)\equiv
        G^E_{\alpha \beta, \mu \nu}(1,2)-        G^E_{\alpha \beta}(1)
G^E_{\mu \nu}(2)
        . \eqn\Cdef $$
is evaluated to be
$$\eqalign{ C_{uu,uu}(1,2) &= {1 \over 8 \pi}
        {U'(1)^2 U'(2)^2 \over (U(2)-U(1))^4}, \cr C_{vv,vv}(1,2) &= {1
\over 8 \pi}
        {1 \over (v(2)-v(1))^4}, \cr C_{uu,vv}(1,2) &= {1 \over 8 \pi}
        {U'(1)^2 \over (v(2)-U(1))^4}. }\eqn\Cexpl $$
Not unexpectedly, the correlations diverge for two points connected by a
light-like line in the direction of the energy flux in question. Note that
there
are correlations between leftward and rightward flux, as anticipated
of this chapter.  The correlations are, however, by no means sharply
localized
\foot{Carlitz and Willey [\carlwil ] claimed
to have found a unique local correlation between
quanta on the two sides. Adapting their  argument to our notation, they
computed \Cexpl\ in the $u$-$v$-frame in the
case where a horizon forms and  analytically continued to
$\tilde v$ defined beyond the  horizon. The
correlation is local in these analytically continued variables,
but not in the
physical ones.}.

One may compare these expressions
to the thermal correlation function (populating
only right-movers) which is easily found to be
$$\eqalign{G_{th}(1,2)=&-{T \over 4} ( |u_1-u_2|+|v_1-v_2| )+ \cr
        &+{1\over 4 \pi} \ln \left[ \left(1-e^{-2 \pi T |u_1-u_2|} \right)
         \left(1-e^{-2 \pi T|v_1-v_2|} \right) \right],}
        \eqn\gthc $$
and to
the two-point
correlation of outward flux:
$$C_{uu,uu}(1,2)={\kappa^4 \over 8 \pi^2}
        {e^{2 \kappa |u_1-u_2|} \over  (e^{\kappa |u_1-u_2|} -1)^4 }~.
        \eqn\twothd $$
There is perfect agreement if we substitute for $U$
the particular trajectory $U \propto -e^{-\kappa u}$.
Moreover, with that choice,
$${\partial \over \partial u_1}{\partial \over \partial u_2}
        G_{th}(1,2)=
        {\partial \over \partial u_1}{\partial \over \partial u_2}
        G_{vac}(1,2),
        \eqn\gtheqgvac $$
so that {\it all\/} correlations of outward energy flux will be thermal.
Correlations involving $T_{vv}$ are, of course, not thermal. In fact, we see
from $\Cexpl$ that the two-point correlation
$C_{vv,vv}(1,2)$ for the mirror is the ordinary correlation expected for a
vacuum state.

In view of our previous qualitative discussion it
is appropriate to note three properties of the radiation:

1.  For the mirror trajectory of the specified form, it is precisely
thermal, in the sense that all its correlation functions are
precisely those one would find for radiation from a black body of
temperature $T$.

2.  It is determined locally from the vacuum state in the
past, in the sense that the correlation
functions in an interval at late times at spatial infinity
can be determined by propagating
a finite interval of vacuum at $t=0$ forward, bouncing once
off the mirror.  In particular,
the correlations in this
interval are not affected by what the mirror does much later.

3.  Nevertheless if, as in Figures 2 and 3, the mirror does not asymptote
to the speed of light, then the final quantum state is pure.
The apparently unlimited entropy of associated with a long period
of thermal radiation is thus, from a microscopic point of view, illusory.
This radiation is not truly thermal; it is correlated with
subsequent``vacuum''
fluctuations.

Evidently it is dangerous to think of microscopic, fine-grained
entropy as a substance which can be measured locally
and once created is never destroyed.  Why one can
get away with this metaphor
for coarse-grained entropy in thermodynamics is
another story ...

\chapter{Limitations and Extensions}

I hope this discussion has
illustrated the use of the moving mirror as a conceptually
transparent model for some aspects of black hole radiance.
Its strength is that it is easy to think about and is a
more-or-less
normal quantum mechanical
system against which one can check alleged black hole
conundrums.  Its obvious weaknesses are that the mirror
is treated as a given rather than being dynamically
determined by the matter, and that the effects of spatial
curvature are ignored.  I would like to close
by mentioning a few directions in which the model
invites extension.

As I mentioned before, to simulate complete evaporation of
a black hole one would like to bring the mirror back
to rest.  An important
limitation to this possibility, however, is that the mirror
may radiate as it is decelerated.  Indeed
if we define
$$\sqrt{U'}=e^{-g},
        \eqn\defg$$
then we obtain from \tmunu\
the total energy flux radiated after the thermal
period in the form:
$$E={1 \over 12 \pi} \int_{u_e}^{u_r} \left( g'^2 -g'' \right) du.
        \eqn\miraa$$
At $u_e$, the end of the thermal period, we have $g \approx \kappa u_e/2$
and $g' \approx \kappa/2$. If we demand that the  mirror be at rest after
$u=u_r$ (so that $g=g'=0$) and minimize the integral \miraa , the $g''$-term
leads to a constant boundary-term in the variational procedure and a
linearly decreasing $g$ is optimal. The trajectory is therefore of
the thermal form \trath\ (with negative $\kappa$).
The integrated flux
decreases with increasing available time.
If we suppose that deceleration sets in only when the hole
has reached the Planck mass, then the available energy is quite
small and one must stretch out the deceleration process in order
to minimize the radiation.
Carlitz and Willey thus deduced
that the time interval over
which the mirror gets back to rest, and  space-time returns to normal,
would have be much longer than the lifetime of the black hole.

While this sort of slowly cooling remnant appears to be a logically
consistent possibility, in the absence of a specific mechanism
it seems a sufficiently strange outcome that one is open to
alternatives.
I find it appealing to imagine that
an appropriate analogue of spatial curvature outside the
hole, which is surely there but is ignored in the simple
mirror model, could
play an essential role in mitigating the problem of radiation
accompanying mirror deceleration.  The analogue would be
a potential barrier to the right of the mirror, whose height
depends on the mass of the acceleration of the mirror (\ie\
mass of the analogue
hole) and becomes very large as the hole mass approaches the
Planck mass.  It is plausible that a mirror hidden behind a
sufficiently high
potential barrier could decelerate to rest in a limited
time without catastrophic emission
of radiation.
If the barrier becomes truly infinite, we have a mirror analogue
of the black hole pinching off into another universe.  In this case
the wave function on {\cal I}$_+$ is insufficient to reconstruct the
original state.  In a sense information is lost, but of course
in the mirror analogue we know exactly where to find it.

Another possible extension is to promote the mirror to a
quantum mechanical variable, instead of a fixed source.  Its state will
then be correlated in a non-trivial way with the emitted radiation.   The
most primitive effect is that it recoils, and thus its position is
described by a wave function correlated with the wave function of
the emitted radiation.   With a heavy mirror treated
semi-classically, this would presumably provide a model for
the intrinsic black hole entropy which appears in classic
black hole thermodynamics.

\refout

\end